%Paper: hep-th/9209003
%From: Bunji Sakita <sakita@sci.ccny.cuny.edu>
%Date: Tue, 1 Sep 92 09:42:34 -0400
%Date (revised): Tue, 8 Sep 92 16:10:50 -0400

%%%%%%%%%%%%%%%%%%%%%% USE PHYZZX FORMAT %%%%%%%%%%%%%%%%%%%%%%%%%%%%%%%%%%%%%%

\input phyzzx

\indent \hfill{CCNY-HEP-92/6}\break
\indent \hfill August 1992 \break

\vskip 0.1 in
{\centerline{\bf FERMIONS IN THE LOWEST LANDAU LEVEL: }}
{\centerline {\bf Bosonization, $W_{\infty}$ Algebra, Droplets, Chiral Bosons}}
\vskip 0.3 in
\centerline {{\bf Satoshi Iso}\foot{e-mail address:
iso@tkyvax.phys.s.u-tokyo.ac.jp}}
\vskip 0.1 in
\centerline {Department of Physics, University of Tokyo, Bunkyo-ku, Tokyo 113,
Japan}
\vskip 0.2 in
\centerline {{\bf Dimitra Karabali}\foot{e-mail address:
karabal@sci.ccny.cuny.edu. Address after September 1, 1992: Syracuse
University,
Physics Department, Syracuse, NY 13244}{\bf and B. Sakita}\foot{e-mail address:
sakita@sci.ccny.cuny.edu}} \vskip 0.1 in
\centerline{Department of Physics, City College of the City University of New
York}
\centerline{New York, NY 10031}
\vskip 0.3 in
\centerline{ABSTRACT}

We present field theoretical descriptions of massless (2+1) dimensional
nonrelativistic fermions
in an external magnetic field, in terms of a fermionic and bosonic second
quantized language.
An infinite dimensional algebra,
$W_{\infty}$, appears as the algebra of unitary transformations which preserve
the lowest Landau level condition and the particle number. In the droplet
approximation it reduces to the algebra of area-preserving diffeomorphisms,
which is responsible for the existence of a
universal chiral boson Lagrangian independent of the electrostatic potential.
We argue that the bosonic droplet approximation is the strong magnetic field
limit of the fermionic theory.
The relation to the $c=1$ string model is discussed.
\pagenumber=0
\eject

\REF\GP{e.g. R. E. Prange and S. M Girvin eds., {\it ``The Quantum Hall Effect"
}, Springer, New York (1990). }
\REF\IKS{S. Iso, D. Karabali and B. Sakita ``One-Dimensional Fermions as
Two-Dimensional Droplets via Chern-Simons Theory", CCNY-HEP-92/1, to appear in
{\it Nucl. Phys. B}.}
\REF\BIPZ{E. Br\'ezin, C. Itzykson, G. Parisi, and J.B. Zuber, {\it Comm. Math.
Phys.} {\bf 59} (1978) 35.}
\REF\DS{E. Br$\acute{\rm e}$zin, V.A. Kazakov and A.B. Zamolodchikov, {\it
Nucl.Phys.} {\bf B338} (1990) 673; G. Parisi, {\it Phys.Lett.} {\bf B238}
(1990)
209; D. Gross and N. Miljkovi$\acute{\rm c}$, {\it Phys.Lett.} {\bf B238}
(1990)
217; P. Ginsparg and J. Zinn-Zustin, {\it Phys.Lett.} {\bf B240} (1990) 333; J.
Polchinski, {\it Nucl.Phys.} { \bf B346} (1990) 253. }
\REF\MO{V. Kazakov, preprint LPTENS 90/30; G. Moore, preprint YCTP-P8-91,
RU-01-12; I. Klebanov, preprint PUPT-1271.}
\REF\TF{e.g. C. Kittel, {\it ``Introduction to Solid State Physics''},
John Wiley and Sons, New York (1986).}
\REF\POL{J. Polchinski, {\it Nucl. Phys.} {\bf B362} (1991) 125.}
\REF\DDMW{S.R. Das, A. Dhar, G. Mandal and S.R. Wadia, ETH, IAS and
Tata preprint, ETH-TH-91/30, IASSNS-HEP-91/52 and TIFR-TH-91/44, to appear
in Int. J. Mod. Phys. A.}
\REF\JS{A. Jevicki and B. Sakita, {\it Nucl. Phys.} {\bf B165} (1980) 511.}
\REF\KS{D. Karabali and B. Sakita {\it Int. J. Mod. Phys.} {\bf A6} (1991)
5079.}
\REF\edge {X. G. Wen, {\it Phys. Rev. Lett.} {\bf 64}(1990) 2206;
M. Stone, {\it Ann. Phys.} (NY) {\bf 207} (1991) 38; J. Fr\"ohlich
and T. Kerler, {\it Nucl. Phys.} {\bf B354} (1991) 365.}
\REF\H{J. Hoppe, MIT Ph. D. Thesis (1982); J. Hoppe and P. Schaller, {\it Phys.
 Lett.}{\bf B237} (1990) 407; C.N. Pope, L.J. Romans and X. Shen
{\it ``A Brief History of $W_{\infty}$''} in Strings 90, ed. R. Arnowitt et al
(World Scientific 1991) and references therein.}
\REF\winf{E. Sezgin, preprint CTP-TAMU-13/92.}
\REF\GMP{M. Girvin, A. H. MacDonald and P. M.
Platzman, {\it Phys. Rev.} {\bf B33} (1986) 2481.}
\REF\WINF{J. Avan and A. Jevicki, {\it Phys. Lett.} {\bf B266} (1991) 35 and
Brown preprint BROWN-HET-824; M. Awada and
S.J. Sin, {\it Phys. Lett.} {\bf B272} (1991) 17; A. Gerasimov, A. Marshakov,
A. Mironov, A. Morozov and A. Orlov, {\it Nucl. Phys.} {\bf B357} (1991) 565;
D. Minic, J. Polchinsky and Z. Yang, Univ. of
Texas preprint UTTG-16-91; G. Moore and N. Seiberg, Rutgers and Yale preprint
RU-91-29, YCTP-P19-91; I. Klebanov and
A.M. Polyakov, Princeton University preprint PUPT-1281; E. Witten
IASSNS-HEP-91/51.}
\REF\ddmw {S.R. Das, A. Dhar, G. Mandal and S.R. Wadia, {\it Mod. Phys. Lett}
{\bf A7} (1992) 71 and 937; A. Dhar, G. Mandal and S.R. Wadia, preprints
IASSNS-HEP-91/89 and TIFR-TH-91/61.}
\REF\GJ{S. M. Girvin and T. Jach, {\it Phys. Rev. } {\bf B29} (1983) 5617.}
\REF\GM{M. Girvin and A. H. MacDonald, {\it Phys. Rev. Lett.} {\bf 58}
(1987) 1252.}
\REF\FJ{R. Floreanini and R. Jackiw, {\it  Phys. Rev. Lett.} {\bf 59} (1987)
1873.}
\REF\SONN{J. Sonnenschein, {\it Nuc. Phys.} {\bf B309} (1988) 752.}
\REF\TAU{C.H. Taubes, {\it Commun. Math. Phys.} {\bf 72} (1980) 277.}
\REF\DMW {A. Dhar, G. Mandal and S. Wadia, preprint TIFR-TH-92/40.}
\REF\CTZ {A. Cappeli, C. Trugenberger and G. Zemba, preprint CERN-TH 6516/92.}
\REF\FS{J. Frohlich and U. Studer ETH-TH preprint.}
\noindent
{\bf Introduction}

The study of nonrelativistic fermions in the presence of electromagnetic field
in (2+1) dimensions is obviously important in condensed matter problems, such
as the quantum Hall
effect (QHE)[\GP ]. Such systems have a further connection with (1+1)
dimensional problems, such as the $c=1$ string model.
In a previous paper [\IKS ] we showed that a system of massless
(2+1) dimensional charged nonrelativistic fermions in a uniform magnetic field
is equivalent to a system
of (1+1) dimensional nonrelativistic fermions, which in an appropriate external
 potential, is
known to describe the $c=1$ string model [\BIPZ,\DS].
In the problem of QHE the applied electrostatic potential is absent
except at the edge of the system but the fermions (electrons)
are mutually interacting through Coulomb force.
On the other hand in the $c=1$ string model the mutual interaction is
absent but the fermions are in a static potential
($A_0 = {1\over 2} ( y^2 - x^2 ) -\mu $ in the double
scaling limit [\MO].)

In [\IKS ], using a classical hydrodynamic formulation,
we arrived at the droplet picture, influenced by the Thomas-Fermi type
classical model [\TF,\POL,\DDMW ]. We established that the dynamics of
droplets is described by an effective collective field Lagrangian [\JS,\KS ].
The results derived are entirely consistent with the theory
of edge excitations in the quantum Hall effect [\edge ].

In this paper we describe the various fermionic and bosonic descriptions of
two-dimensional
massless nonrelativistic fermions in an external magnetic field and discuss
the relation between them.

We further study the symmetries of the original system
and how they manifest themselves in both the fermionic and bosonic field
theoretic description. In particular we find the existence of
an infinite dimensional algebra,
$W_{\infty}$ algebra [\H ,\winf  ], which was
noticed in the study of QHE [\GMP] and independently
realized and studied extensively in the framework of $c=1$ string model
[\DDMW,\WINF,\ddmw].
In both the fermionic and bosonic description we find that the $W_{\infty}$
algebra is the algebra of unitary transformations
which preserve the lowest Landau level condition and the particle number.

In the droplet approximation the $W_{\infty}$ algebra is reduced to the
algebra of area-preserving
diffeomorphisms. As a consequence we find that the dynamics of the droplets is
described by a universal chiral boson Lagrangian independent of the details
of the electrostatic potential.

Finally we discuss the relation of our results to the $c=1$ string model.

\noindent{\bf Field Theory of Massless Fermions in External Magnetic Field}

The many-body Hamiltonian of a system of massive
(mass $m$) fermions in electromagnetic
field in two space dimensions is given by
$$\eqalignno{
H&={1\over{2m}}\sum_{a=1} ^N ( {\bf \Pi }_a )^2 + \sum_a A_0
({\bf x}_a )\cr &={1\over{2m}}\sum_{a=1} ^N ( \Pi^x _a +i\Pi^y _a )
( \Pi^x _a -i\Pi^y _a )+ \sum_a A_0 ({\bf x}_a ) +{B\over{2m}}N , &(1)\cr}
$$
where
$$
\Pi^i =  p_i -A^i ({\bf x}) =-i {\partial \over \partial x^{i}}-A^i ({\bf x}),
\ \ \ \ \  \ \ \ \  i= x, y \eqno (2)
$$
and $B$ is a uniform external magnetic field defined by $\vec{\nabla} \times
\vec {A}=-B$. In what follows we shall omit the last constant term in (1).
The corresponding Schr\"odinger wave function
$\Psi ( {\bf x}_1 , {\bf x}_2 , \cdots {\bf x}_N
)$ is a totally antisymmetric function of
${\bf x}$'s. As is well known, the fermions occupy discrete,
degenerate Landau levels,
uniformly separated by an energy gap $E_c = {B \over m}$
(we have set $\hbar=c=e=1$).

The massless limit of the above system restricts the fermions to the lowest
Landau level for which the wavefunction obeys
$$
( \Pi^x _a -i\Pi^y _a ) \Psi ( {\bf x}_1 , {\bf x}_2 ,
\cdots {\bf x}_N )=0\eqno (3)
$$
In the symmetric gauge, where $A^{x}= By/2$, $A^{y}= -Bx/2$, the lowest
Landau level condition is written as
$$
(\partial _{z_{a}} + {{\bar{z}_{a}} \over 2}) \Psi ({\bf x}_1 , {\bf x}_2 ,
\cdots {\bf x}_N )=0
\eqno(4)
$$
where $z_{a} = \sqrt{{B \over 2}}(x_{a}+iy_{a})$,
$\bar{z_{a}} = \sqrt{{B \over 2}}(x_{a}-iy_{a})$. The solution to (4) is given
by
$$ \Psi ( {\bf x}_1 , {\bf x}_2 , \cdots {\bf x}_N
 ) = f(\bar{z}) \exp ({-{\textstyle {1 \over 2}}\sum _{a=1}
^{N} |z_{a}| ^2}) \eqno (5)
$$
where $f(\bar{z})$ is a polynomial in the variables $\bar{z_{a}}$.

The projection of the Hamiltonian $H=A_0 (\hat{z}_a, \hat{\bar{z}}_a)$
\footnote{*}{ We consistently use $f(z,\bar{z})$ for
$f(x,y) \big|_{{{x=\sqrt{{1\over{2B}}}(z+\bar{z})}\atop
{y={1\over i}\sqrt{{1\over{2B}}}(z-\bar{z})}}}$} onto
the lowest Landau level is given by
$$  H = \sum_a \ddag A_0 ({\partial \over \partial \bar{z}_a}+{z_a \over 2},
\bar{z}_a ) \ddag \eqno (6) $$
where $\ddag ~~~\ddag$ denotes the ordering where all the
derivatives are on the left [\GJ].

One can bosonize this system by using a
singular gauge transformation [\GM ].
$$
\Psi ( {\bf x}_1 , {\bf x}_2 , \cdots {\bf x}_N)
= e^{i \sum_{a>b} Im \ln (\bar{z}_a - \bar{z}_b )}\Phi ( {\bf x}_1 , {\bf x}_2
,
\cdots {\bf x}_N) \eqno (7)
$$
where $\Phi ( {\bf x}_1 , {\bf x}_2 , \cdots {\bf x}_N ) $
is a symmetric wave function.  As a result we obtain, for the massless case, an
effective bosonic Hamiltonian which is of the form
$$  H = \sum_a \ddag A_0 ({\partial \over \partial \bar{z}_a}+{z_a \over 2}
+{1 \over 2} \sum _{b \ne a} {1 \over {\bar{z}_a - \bar{z}_{b}}},
\bar{z}_a ) \ddag \eqno (8) $$
 The bosonized analog of
the lowest Landau level condition (4) takes the form
$$
( {\partial \over \partial z_a}+{\bar{z}_a \over 2}
-{1 \over 2} \sum _{b \ne a} {1 \over {z_a- z_b}}) \Phi ( {\bf x}_1 ,
{\bf x}_2 , \cdots {\bf x}_N )=0\eqno (9)
$$

The massless quantum mechanical system given in (3)-(9) can be further
expressed in a second quantized
language. There are two equivalent field theory
descriptions for the fermionic representation
(3)-(6).

The first description (``unreduced")
is to introduce the standard fermion creation and
annihilation operators
$$
\{ \hat \psi ({\bf x}) ,\hat\psi^{\dag}({\bf x}')\}=\delta ({\bf x}-{\bf x}')
\eqno (10)
$$ and define
$$
\Psi ({\bf x}_1 , {\bf x}_2 , \cdots {\bf x}_N ) ={1\over{\sqrt{N!}}}\bra{0}
\hat \psi ({\bf x}_1 )\hat \psi ({\bf x}_2 )\cdots \hat \psi ({\bf x}_N )\ket
{\Psi}\eqno (11)
$$
The lowest Landau level condition (4) and the Hamiltonian (6)
are derived respectively from
$$
(\partial _{z} + {{\bar{z}} \over 2})\hat{\psi}({\bf x})\ket{ \Psi }=0
\eqno (12)
$$
$$
\eqalign{ H &=\int d {\bf x}\hat{\psi }^{\dag}({\bf x})
\ddag A_0 ({\partial \over \partial \bar{z}}+{z \over 2},
\bar{z} ) \ddag \hat{\psi}({\bf x})\cr
&\ \dot{=}\int d {\bf x}\hat{\psi }^{\dag}({\bf x})
 A_0 ({\bf x}) \hat{\psi}({\bf x})\cr}\eqno (13)
$$
where $\dot{=}$ is an equality up to an operator which vanishes when it acts
on $\bra{\Psi}$ from the right.

The second description (``reduced") is to impose the operator constraint
$$
\left( \partial_z +{1\over 2}\bar{z}\right)\hat{\psi} ({\bf x},t)=0\eqno (14)
$$
which again produces the lowest Landau level condition (4).
A general solution of (14) has the form
$$
\hat{\psi} ({\bf x},t) = {\sqrt{B\over {2\pi}}}\ e^{-{1\over 2}|z|^2 }
\sum_{n=0} ^{\infty}{{\bar{z} ^n} \over {{\sqrt {n!}}}}
\hat{C}_n (t)    \eqno (15)
$$
where the modes $\hat{C}_n$ satisfy the usual anticommutation relations
$\{\hat{C}_n, \hat{C}^{\dag} _m
\}=\delta _{nm}$. The constrained $\hat{\psi}$ operators no longer satisfy the
commutation relation (10). In terms of the constrained operators
the Hamiltonian is given by (13) and is exactly equal to the expression in the
second line of (13).

It can be easily shown that these two descriptions are equivalent.
We only remark that they can be considered
as two different methods of quantizing the
following classical Lagrangian
$$
{\cal L} = \bar{\psi}({\bf x},t)\left( i{{\partial}\over{\partial t}}
- A_0 ({\bf x} )\right) \psi ({\bf x},t) \eqno (16)
$$
with the constraint
$\left( \partial_z +{1\over 2}\bar{z}\right){\psi} ({\bf x},t)=0$.
The reduced method amounts to solving the constraint explicitly, substituting
the solution into the Lagrangian and developing a canonical quantization for
the unconstrained variables.

Similarly the bosonized massless quantum mechanical system
(7)-(8) is expressed, in a second quantized
language, in terms of a Hamiltonian given by
$$\eqalignno{ H &= \int d{\bf x}  \hat {\phi} ^{\dag}({\bf x})
\ddag A_0 ({\partial \over {\partial \bar{z}}}+{z \over 2 }+{1 \over 2} \int
d{\bf x}'{{\hat{\rho}({\bf x}')}
\over {\bar{z}-\bar{z}'}},
\bar{z} )\ddag  \hat{\phi} ({\bf x}) \cr
&\ \dot{=}
\int d{\bf x}  \hat {\phi} ^{\dag}({\bf x})
A_0 ({\bf x} ) \hat{\phi} ({\bf x}) &(17) \cr}$$
where $\hat\phi$ is a bosonic field operator
satisfying the usual commutation relations
$[\hat{\phi}(\bf{x}), \hat{\phi}^{\dag}(\bf{x'})]=\delta (\bf{x}-\bf{x'})$ and
$\hat{\rho}({\bf x})
\equiv \hat{\phi}
^{\dag} ({\bf x}) \hat{\phi} ({\bf x})$.
The analog of the lowest Landau level condition is of the form
$$
\left({\partial \over \partial z}+{\bar{z} \over 2 }-{1 \over 2} \int {
{\hat{\rho}({\bf x}')}
\over {z-z'}}\right)\hat{ \phi} ({\bf x},t)\ket{\Phi}=0\eqno (18)
$$
This is the bosonic analog of the unreduced description.

We should remark here that in the bosonic formalism the lowest Landau level
condition (18) cannot be straightforwardly reduced to
an operator equation because the constraint equation is nonlinear in
$\hat{\phi}$.

If one ignores the subtlety of operator ordering, namely
the order $\hbar$ corrections, it is posssible to describe the system
in terms of a classical Lagrangian of the same form as in (16), where $\psi$
is replaced by $\phi$, with the constraint
$$
\left({\partial \over \partial z}+{\bar{z} \over 2 }-{1 \over 2} \int
{{\rho({\bf x}'
)}
\over {z-z'}}\right) \phi ({\bf x},t)=0 \eqno (19)
$$

In terms of the reduced formalism one would solve the classical constraint
equation (19) and then develop a canonical formalism for the unconstrained
variables. A priori this may
not be equivalent to the bosonic representation described by (17)-(18) at
the quantum mechanical
level, but presumably a proper choice of operator ordering in terms of the
unconstrained variables
can be made so that the fully quantum mechanical bosonic representation is
recovered.

In [2] the bosonic problem was further formulated in terms of
hydrodynamic variables
$$
{\phi =\sqrt{\rho} e ^{i\theta}} \eqno (20)$$
It was shown that the corresponding Lagrangian has the form
$$
L= \int d^{2}x d^{2}x' \rho (x) 2 \pi G(x-x')
\dot{\rho}_V (x') - \int d^2 x \rho (x) A_0 \eqno (21) $$
while the classical lowest Landau level condition is written as
$$
B + {1\over 2}\nabla^2\ln\rho -2\pi\rho_V
-2\pi\rho =0\eqno (22)
$$
where $ \rho _{V} ({\bf x}) $ denotes the vortex distribution function
defined by
$$-2 \pi \rho _{V} ({\bf x})=\epsilon^{ij} \partial_i \partial_j \theta
({\bf x}) \eqno (23)
$$
and $G$ is the Green's function
satisfying
$$
{\epsilon^{ij}\partial_i\partial_j G(x-x') = \delta^2 (x-x'),
\ \ \ \ \ \ \ \ \nabla^2 G=0}.\eqno (24)
$$

One can show that (22) provides a way of writing (19) in terms of the
hydrodynamic variables (20).

Since the full quantum mechanical constraint (18) is rather
difficult to implement,
our approach in [2] was to treat (22)
as a classical constraint equation, solve it and then
develop a canonical formalism for the unconstrained variables.

\noindent
{\bf Two-Dimensional Fermions and $W_{\infty}$ Algebra}

In terms of the standard coherent state representation:
$$
\ket{z} = {e^{z {\hat{a}}^{\dag}}} \ket{0} , \ \ \ \  \int
d^2 z \  {e^{-|z|^2}}
\ket{z}\bra{z}=1 , \ \ {\rm etc.} \eqno (25)
$$
the constrained operator $\hat{\psi}$ in (15) can be written as
$$
\hat{\psi} (x,y) = {\sqrt{B\over {2\pi}}}\ e^{-{1\over 2}|z|^2 }\sum_n
\langle{z}\ket{n} \hat{C}_n\eqno (26)
$$

Next let us consider a unitary transformation in the space of $\hat{C}_n$:
$$
\hat{C}'_n = u_{nm} \hat{C}_m =\bra{n} \hat{u}\ket{m} \hat{C}_m \eqno (27)
$$
An infinitesimal transformation
is generated by a hermititian operator which we write as
$\ddag \xi ( \hat {a} , {{\hat{a}}^{\dag}} )\ddag$ with the anti-normal order
symbol, where $\xi$ is a real function when $\hat{a}$ and ${\hat{a}}^{\dag}$
are replaced by $
z$ and $\bar z$ respectively. Then using (26) we obtain the following
infinitesimal transformation for $\hat{\psi}$:
$$
\delta \hat{\psi} (x,y) =i{\sqrt{B\over {2\pi}}}\ e^{-{1\over 2}|z|^2 }
\ddag \xi ( \partial_{\bar{z}} , \bar{z} )\ddag \sum_n \langle{z}\ket{n}
\hat{C}_n
=i\ddag \xi ( \partial_{\bar{z}}+{1\over 2}z , \bar{z} )\ddag \hat{\psi} (x,y)
\eqno (28)
$$
where $\ddag\ \ \ \ddag$ indicates that
the derivatives are
placed on the left of
$z$ and $\bar {z}$.
The charge density $\hat{\rho} = \hat{\psi}^{\dag}\hat{\psi} $ is transformed
as
$$
\delta \hat{\rho} (x, y) = i\left( \ddag \xi ( \partial_{\bar{z}}+ z ,
\bar{z} )\ddag - \ddag\xi ( z , \partial_{{z}} + \bar{z} )\ddag\right )
\hat{\rho} (x,y)\eqno (29)
$$
As a result we find, by partially integrating (29), that the total fermion
number remains invariant as it should be:
 $$\int dx dy \delta\hat{\rho} (x,y) =0\eqno (30)$$

Since $\partial_z +\bar{z}/2$ commutes with $\partial_{\bar{z}}+z/2$
the transformation (28) can be considered as the most general linear
infinitesimal
transformation which preserves the lowest Landau level condition (14). It is
straightforward to find that the generator of
this transformation is given by
$$
\hat{\rho}[{\xi}] \equiv \int dx dy \xi (x,y)\hat{\rho} (x,y) =\int dx dy
\hat{\psi} ^{\dag} (x,y)
\ddag\xi ( \partial_{\bar{z}}+{1\over 2}z , \bar{z} )\ddag \hat{\psi} (x,y)
\eqno (31) $$
Using (15) and the anticommutation relation for $\hat{C}$'s we find that the
algebra satisfied by
the generators $\hat{\rho}[{\xi}]$ is given by
$$
[\,\hat{\rho} [\,\xi_1 \, ],\hat{\rho} [\, \xi_2 \, ]\, ]={i\over B}\hat{\rho}
[\{\!\!\{\xi_1 ,\xi_2 \}\!\!\}] \eqno (32)
$$
where
$$
\{\!\!\{\xi_1 ,\xi_2 \}\!\!\}=iB{\sum_{n=1} ^{\infty}}{{(-)^n}\over{n!}}\left(
{\partial_{z} ^{n}}\xi_1 {\partial_{\bar{z}} ^n}\xi_2 -
{\partial_{\bar{z}} ^{n}}\xi_1 {\partial_{z} ^n}\xi_2\right)\eqno (33)
$$

By choosing $\xi (x,y) = \exp i(px +qy)$ we obtain the
commutation relation of
the fermion density in the momentum space:
$$ [\hat{\rho} (p,q), \hat{\rho} (p',q')]=-2i \sin ({{pq'-qp'} \over 2B})
e^{{{pp'+qq'} \over 2B}} \hat{\rho} (p+p',q+q') \eqno (34) $$
where $\hat{\rho} (p,q)= \int dxdy e^{ipx} e^{iqy} \hat{\rho} (x,y) \equiv
\hat{\rho} [\exp i(px +qy)]$.
And also by choosing $\xi (z, \bar{z}) =z^l \bar{z}^m $ we obtain
$$[\hat{\rho}_{rs}, \hat{\rho}_{lm}]= \sum_{n=1}^{min(l,s)} {{(-1)^{n}}
\over { n!}}
{{l!s!} \over {(l-n)!(s-n)!}} \hat{\rho}_{r+l-n, s+m-n}
-( s \leftrightarrow m, l \leftrightarrow r)       \eqno (35) $$
where $\hat{\rho}_{lm} =\int dx dy z^l \bar{z}^m\hat{\rho}
\equiv\hat{\rho}[z^l \bar{z}^m ]$, which is a
coefficient of the power series expansion of $\hat{\rho} (p,q)$.
The Lie algebra (32) and its representations (34) and (35)
in the specific bases are manifestations of the $W_{\infty}$ algebra [\H ],
which in this case is the algebra of
$U({\infty})$. It corresponds to unitary transformations which preserve the
lowest Landau level condition and the fermion number.

The Hamiltonian, given by $ H=\int A_0\hat{\rho}$, is also an element of
$W_{\infty}$. Using the Heisenberg equation of motion and the commutation
relation (32) we obtain the equation of motion
for the fermion density projected on the
lowest Landau level:
$$\partial_{t} \hat{\rho} (x,y,t) = i \sum _{n=1}^{\infty}
{1 \over n!} [\partial_{z}^{n}(\hat{\rho}
\partial_{\bar{z}} ^{n} A_0 ) - \partial_{\bar{z}}^{n}
( \hat{\rho} \partial_{z} ^{n} A_0 )] \eqno(36)$$

We now turn to the bosonic formalism (17)-(18).
Let us look for a unitary transformation on the state $\ket{\Phi}$ which
preserves (18). This is
generated by an operator ${\cal{O}}$ such that
$$
[(\partial_{z} + {\bar{z} \over 2}- {1 \over 2} \int d{\bf x}' {{\hat{\rho}
({\bf x}')} \over {z-z'}})
\hat{\phi}({\bf x}), ~ {\cal{O}}] \ket{\Phi} =0 \eqno(37)
$$
We identify the operator ${\cal{O}}$ to be
$${\cal{O}}[{\xi}] = \int d{\bf x} \hat{\phi} ^{\dag} ({\bf x}) \ddag \xi
(\partial_{\bar{z}} +
{z \over 2}+
{1 \over 2} \int d{\bf x}' {{\rho({\bf x}')} \over {\bar{z}-\bar{z}'}},
\bar{z}) \ddag \hat{\phi} ({\bf x})
\eqno (38)
$$
One can now show using (37) that the operator ${\cal{O}}[{\xi}]$ generates
the $W_{\infty}$
algebra on the space of states satisfying the lowest Landau level condition,
namely
$$
\bra{\Phi}  [{\cal{O}}[{\xi_{1}}], {\cal{O}}[{\xi_{2}}]] = <\Phi |{i \over B}
{\cal{O}}[{\{\!\!\{\xi_{1},
\xi_{2}\}\!\!\}}
] \eqno (39) $$
where $\{\!\!\{\xi_{1},\xi_{2}\}\!\!\}$ is defined as in (33).
The Hamiltonian (17) is again an element
of $W_{\infty}$.

We find that in both the fermionic and bosonic formulation the
$W_{\infty}$-algebra emerges
as the algebra of unitary transformations of physical states.
\eject

\noindent
{\bf Droplet Approximation, Chiral Bosons and $w_{\infty}$ algebra}

The droplet approximation considered in [2] is based on the hydrodynamic
formulation obtained in (21)-(24). In this case an approximate solution to
the lowest Landau level equation (22) is considered, where the particle
density is maximum, $\rho = B/2 \pi$, inside and zero outside a boundary
in space and correspondingly the vortex density is maximum, $\rho_V =B/2 \pi$,
outside and zero inside the boundary.
The only dynamical variable in this picture is the one parametrizing
the boundary. The classical ground state configuration corresponds to a
boundary specified by the curve $A_0 (x,y) =0$,
while excited states of the system correspond to deformations of this boundary.

The above choice for $\rho$ and $\rho_{V}$ satisfies the condition (which can
be considered
as the lowest Landau level condition in the droplet approximation)
$$
2\pi\rho + 2\pi\rho_V -B=0,\ \ \ \ \ \rho \, \rho_V =0\eqno (40)
$$
and is such that the energy $\int d^2 x A_0 \rho $ stays in the neighborhood of
minimum, namely the
particle density is maximum in the region of negative $A_{0}$ (we assume that
this is the region inside the boundary) and minimum in
the region of positive $A_{0}$ (outside the boundary).

In order to specify the boundary we choose a coordinate system $r,~s$, where
$$
r=A_0 (x,y),\ \ \ \ \ s=S(x,y)
,\ \ \ \ {\rm such\ that}\ \ \ \  dxdy = dr ds\eqno (41)
$$
The transformation (41) is a transformation that preserves
the local area and satisfies
$\{A_0 , S\}_{PB} \equiv \partial_x A_0 \partial_y S-\partial_x S \partial_y
A_0
=1$.
Assuming that the boundary is closed, we parametrize $\rho$ and $\rho_V$ as
$$
 \rho (x,y,t) ={B\over{2\pi}}\theta (r(s,t)-r),
\ \ \ \ \rho_V (x,y,t)={B\over{2\pi}}\theta (r-r(s,t)) \eqno (42)
$$ and regard $r(s,t)$ as a set of dynamical variables.
The above ansatz solves explicitly the constraint equation (40).
We restrict $r(s,t)$ to be a singlevalued function of $s$.
Then the Lagrangian of the system, eq. (21),
is given by
$$
L={B^2 \over {8 \pi}} \oint ds\oint ds' r(s) \epsilon (s-s')\dot{r}(s')
 -{B \over {4 \pi}}\oint ds  r^2 (s)
\eqno (43)
$$
After rescaling the field, namely $r(s,t) \rightarrow \sqrt {2 \pi} r(Bs, t)$,
it reduces to the chiral boson Lagrangian\footnote{*}{
We should remark here that the transformation (41) may have singular points.
This is so if the inverse transformation is multivalued. If this is the case
there exist multi-boundaries and we have to introduce multi-chiral boson
fields, one for each boundary. We ignore this complication in this paper.}
[\FJ ,\SONN ].
The equation of motion for the field $r(s,t)$ is
$$(\partial_{t}-{\textstyle {1 \over B}} \partial_{s})r(s,t) =0 \eqno (44) $$

Based on (43) we quantize the system and we study the corresponding quantum
collective motion of the original fermions as the  quantum mechanical droplet
motion. The canonical commutation relation for the chiral boson field is
$$
[ \hat{r}(s,t) , \hat{r}(s',t)] = {{2 \pi i} \over B^2}\delta ' (s - s') \eqno
(45)
$$

Using (43) and (45) we can show that the density operator $\hat{\rho}$
satisfies the following commutation relation:
$$
[\,\hat{\rho} [\,\xi_1 \, ],\hat{\rho} [\, \xi_2 \, ]\, ]={i\over B}\hat{\rho}
[\{\xi_1 ,\xi_2 \}_{PB}] \eqno (46)
$$
where
$$
\hat{\rho} [\,\xi \, ] =\int dx dy \xi (x,y) \hat{\rho} (x,y) \ \ \ {\rm and}
\ \ \{\xi_1 ,\xi_2 \}_{PB}=
\epsilon_{ij}\partial_{i}\xi_1 \partial_{j}
\xi_2\eqno (47)
$$
and $\xi (x, y )$ is a real function.
This is the algebra of area-preserving diffeomorphisms, which is also
called classical $w_{\infty}$ algebra. Thus the $\hat{\rho} $'s generate the
area-preserving diffeomorphisms as
quantum mechanical unitary transformations. It is obvious that these
transformations preserve
the particle number, or in other words the area of the droplet since
$$
\delta _{\xi} \hat{\rho} (x,y) = {1 \over B} \epsilon_{ij}\partial_{i}\xi
\partial_{j} \hat{\rho} (x,y) \eqno(48)
$$

Since the Hamiltonian of the system is given by
$$
H=\int dx dy A_0 (x,y)\hat{\rho} (x,y) \equiv \hat{\rho} [A_0 ] \eqno (49)
$$
it is an element of the algebra.  Therefore all the systems
with $A_0$'s related by area-preserving transformations
are unitarily equivalent.

Let us now discuss the relationship between
the full fermionic theory and
the droplet approximation we used for the corresponding bosonic theory.

Let us first compare the algebras (32)-(33) and (46)-(47).
Since $z$ is proportional to $\sqrt B$, in the large $B$ limit one may
neglect the higher derivative terms in (33). In this limit the $W_{\infty}$
algebra is reduced to the classical $w_{\infty}$ algebra:
$$
\{\!\!\{ \xi_{1}, \xi_{2} \}\!\!\} \rightarrow  \{ \xi_{1}, \xi_{2} \}_{PB}
\eqno(50)
$$
More generally we can say that the droplet approximation is the large $B$
limit of the hydrodynamic formulation described by (21)-(24), which is
equivalent to the original
fermion theory at least semiclassically. In order to justify this let us go
back to the lowest Landau level condition as expressed in (22). This is the
equation for vortices [\TAU].
$B$ is the only dimensional parameter $B \sim 1/l^2$. If we were to solve
(22) the existence of the term $\nabla^2\ln\rho$ would result to the
softening of the step function ansatz for $\rho (x,y)$ producing a thick
boundary, where the thickness is necessarily of
order $\sim 1/\sqrt{B}$. In that sense the droplet approximation, where the
boundary is sharp, is valid only when $1/\sqrt{B}$ is much smaller than the
size of the separation of
the boundaries of the droplet which is determined by $A_0$.

\eject

\noindent
{\bf Relation to $c=1$ string theory}

As we have mentioned earlier, the system of two-dimensional nonrelativistic
massless fermions in a
uniform magnetic field is equivalent to one-dimensional nonrelativistic
fermions. If we further choose the electrostatic potential to be of the form
$A_0 = {1 \over 2}(y^2 -x^2)-\mu$,
it describes the $c=1$ string model [\MO]. Therefore it is of no surprise
that a similar algebraic
structure, like the $W_{\infty}$ algebra, emerges in both the (2+1)
dimensional fermionic
system and the (1+1) dimensional one which describes the $c=1$ string model.

In a series of papers [\DDMW,\ddmw] the field $W(p,q,t)$ was introduced,
which is a bilinear in terms of one-dimensional fermions and carries the $W$
algebra. In our case it is the two-dimensional fermion density that carries
the $W$ algebra. Due to the lowest Landau level condition one can write the
two-dimensional fermion density in the momentum space as
$$\hat{\rho} (p,q)= \sum_{nl} C_{l} ^{\dag} C_{n}\bra{l}\exp{{i(p-iq)\hat{a}}
\over {\sqrt{2B}}} \exp{{i(p+iq) \hat{a}^{\dag}} \over
{\sqrt{2B}}}\ket{n} \eqno(51) $$
Using this we find that there is a relationship
between the one-dimensional
bilinear $W$ and our two-dimensional fermion density.
$$ \hat{\rho} (p,q,t) = 2 \exp(-{{p^2 +q^2} \over 4B}) W(p,-q,t) \eqno(52) $$
where $B$ plays the role of $\hbar ^{-2}$ in the definition of $W$. This
relation is further extended in deriving
the equation of motion and Ward-identities for the correlators of the
bilinears.

Let us now discuss the relation between the droplet approximation and the
$c=1$ string model. We
argued in the previous section that the droplet approximation is valid only
when $1/\sqrt{B}$ is much smaller than the characteristic length introduced
by $A_0$. Since the potential
appropriate for $c=1$ string model is of the form $A_0 = {1 \over 2}(y^2 -x^2)-
\mu$, the droplet
approximation makes sense when $1/\mu \ll B$, which corresponds to a weak
string coupling $(\mu \rightarrow \infty)$. In order to describe the theory
away from the weak coupling region,
one should solve the constraint (22). In that case the boundary becomes
blurry and the droplet picture disappears as $\mu \sim 1/B$.

\noindent
\centerline{\bf Acknowledgements}

This work was supported by the NSF grant PHY90-20495 and the Professional
Staff Congress
Board of Higher Education of the City University of New York under grant
no. 6-63351.

\vskip 0.5 in

While this work was in progress we received [\DMW],[\CTZ],[\FS] where issues
related to the ones presented here have been discussed in the framework of
the Quantum Hall effect and the $c=1$ string model.
\refout
\end